\documentclass[twocolumn,superscriptaddress,amsfont,amssymb,amsmath, showpacs,balancelastpage, nofootinbib, preprintnumbers]{revtex4-1}
\usepackage{graphicx,longtable,mathrsfs,color,array}
\usepackage[unicode=true,pdfusetitle,
bookmarks=true,bookmarksnumbered=true,bookmarksopen=true,bookmarksopenlevel=1,
breaklinks=false,pdfborder={0 0 0},backref=false,colorlinks=true]{hyperref}
\hypersetup{citecolor=blue,filecolor=blue,linkcolor=blue,urlcolor=blue}
\usepackage[usenames,dvipsnames]{xcolor}
\usepackage{amssymb,amsmath,mathtools,mathrsfs,enumitem}
\usepackage{epsfig,subfigure,placeins,float}
\usepackage{booktabs,longtable,multirow}
\usepackage{exscale,relsize}
\usepackage[normalem]{ulem}
\usepackage[T1]{fontenc}
\usepackage[utf8]{inputenc}
\usepackage{enumerate}
\usepackage{times, mathptmx}

\usepackage{tikz}
\usepackage[compat=1.1.0]{tikz-feynman}
\usetikzlibrary{arrows,decorations.markings}
\usepackage{soul}

\usetikzlibrary{decorations.pathmorphing}

\usetikzlibrary{arrows,positioning,decorations.markings,decorations.pathmorphing,calc}
\definecolor{hyperref}{RGB}{026,028,087}

\usepackage{tikz}
\usepackage[compat=1.1.0]{tikz-feynman}

\newcommand{\gravh}{
\begin{tikzpicture}[baseline]
\begin{feynman}
\vertex (a);
\vertex [right=1cm of a] (b);
\diagram* {
(a) -- [photon] (b)
}; 
\end{feynman}
\end{tikzpicture}
}

\newcommand{\gravH}{
\begin{tikzpicture}[baseline]
\begin{feynman}
\vertex (a);
\vertex [right=1cm of a] (b);
\diagram* {
(a) -- [plain, insertion={0}] (b)
}; 
\end{feynman}
\end{tikzpicture}
}

\newcommand{\Csoure}{
\begin{tikzpicture}[baseline]
\begin{feynman}
\vertex [dot, minimum size=0.25cm] (a) {};
\vertex [right=0.82cm of a] (d);
\vertex [right=0.2cm of d] (e);
\diagram* {
(a) -- [photon] (d),
(d) -- [plain, -stealth] (e)
}; 
\end{feynman}
\end{tikzpicture}
}

\newcommand{\proph}{
\begin{tikzpicture}[baseline]
\begin{feynman}
\vertex [label=90:$\scriptstyle{\mu\nu}$](a);
\vertex [right=1.5cm of a, label=90:$\scriptstyle{\rho\sigma}$] (b);
\diagram* {
(a) -- [photon, momentum={[arrow shorten=0.3pt]$k$}] (b)
}; 
\end{feynman}
\end{tikzpicture}
}

\newcommand{\cubicback}{
\begin{tikzpicture}[baseline]
\begin{feynman}
\vertex (a);
\vertex [right=1cm of a] (b);
\vertex [right=1cm of b] (c);
\vertex [above=0.5cm of c, label=90:$\scriptstyle{\alpha_2\beta_2}$] (du);
\vertex [below=0.5cm of c, label=270:$\scriptstyle{\alpha_3\beta_3}$] (dd);
\diagram* {
(a) -- [plain, insertion={0}] (b),
(du) -- [photon, momentum'={[arrow shorten=0.3pt]$k_2$}] (b),
(dd) -- [photon, momentum={[arrow shorten=0.3pt]$k_3$}] (b)
}; 
\end{feynman}
\end{tikzpicture}
}

\newcommand{\quarticback}{
\begin{tikzpicture}[baseline]
\begin{feynman}
\vertex (a);
\vertex [above=0.5cm of a] (Hu);
\vertex [below=0.5cm of a] (Hd);
\vertex [right=1cm of a] (b);
\vertex [right=1cm of b] (c);
\vertex [above=0.5cm of c, label=90:$\scriptstyle{\alpha_3\beta_3}$] (du);
\vertex [below=0.5cm of c, label=270:$\scriptstyle{\alpha_4\beta_4}$] (dd);
\diagram* {
(Hu) -- [plain, insertion={0}] (b),
(Hd) -- [plain, insertion={0}] (b),
(du) -- [photon, momentum'={[arrow shorten=0.3pt]$k_2$}] (b),
(dd) -- [photon, momentum={[arrow shorten=0.3pt]$k_3$}] (b)
}; 
\end{feynman}
\end{tikzpicture}
}

\newcommand{\cubic}{
\begin{tikzpicture}[baseline]
\begin{feynman}
\vertex [label=180:$\scriptstyle{\alpha_1\beta_2}$](a);
\vertex [right=1cm of a] (b);
\vertex [right=1cm of b] (c);
\vertex [above=0.5cm of c, label=90:$\scriptstyle{\alpha_2\beta_2}$] (du);
\vertex [below=0.5cm of c, label=270:$\scriptstyle{\alpha_3\beta_3}$] (dd);
\diagram* {
(a) -- [photon, momentum={[arrow shorten=0.3pt]$k_1$}] (b),
(du) -- [photon, momentum'={[arrow shorten=0.3pt]$k_2$}] (b),
(dd) -- [photon, momentum={[arrow shorten=0.3pt]$k_3$}] (b)
}; 
\end{feynman}
\end{tikzpicture}
}

\newcommand{\Source}{
\begin{tikzpicture}[baseline]
\begin{feynman}
\vertex [dot, minimum size=0.25cm, label=180:$M$] (a) {};
\vertex [right=1cm of a] (d);
\vertex [right=0.2cm of d, label=90:$\scriptstyle{\mu\nu}$] (f);
\diagram* {
(a) -- [photon, momentum={[arrow shorten=0.3pt]$k$}] (d),
(d) -- [plain, -stealth] (f)
}; 
\end{feynman}
\end{tikzpicture}
}

\newcommand{\Ps}{\mathbb{P}}

\def\gsim{ \lower .75ex \hbox{$\sim$} \llap{\raise .27ex \hbox{$>$}} }
\def\lsim{ \lower .75ex \hbox{$\sim$} \llap{\raise .27ex \hbox{$<$}} }
\def\be{\begin{equation}}
\def\ee{\end{equation}}
\def\bea{\begin{eqnarray}}
\def\eea{\end{eqnarray}}

\DeclareMathOperator{\e}{e}
\newcommand{\gEFT}{g^{\rm EFT}}
\newcommand{\gbRW}{g^{\rm Sch}}
\newcommand{\MPl}{M_{\rm Pl}}

\newcommand{\comment}[1]{}
\newcommand{\Mpl}{M_{\rm Pl}}

\newcommand{\hypergeom}[2]{
	  \mathbin{_{#1}{\sf F}_{#2}} }

\def\D{{\rm d}}
\def\dd{{\rm d}}

\definecolor{Gray}{gray}{0.9}
\definecolor{LightCyan}{rgb}{0.88,1,1}

\renewcommand*{\mathcolor}{}
\def\mathcolor#1#{\mathcoloraux{#1}}
\newcommand*{\mathcoloraux}[3]{%
\protect\leavevmode
\begingroup
\color#1{#2}#3%
\endgroup
}

\newcommand{\textins}[2][fu-grey]{
\ifmmode\mathcolor{#1}{#2}
\else\textcolor{#1}{#2}\@\,
\fi
}
\graphicspath{{./}}
\allowdisplaybreaks

\tikzstyle{vecArrow} = [thick, decoration={markings,mark=at position
1 with {\arrow[semithick]{open triangle 60}}},
double distance=1.4pt, shorten >= 5.5pt,
preaction = {decorate},
postaction = {draw,line width=1.4pt, white,shorten >= 4.5pt}]
\begin{document}


\preprint{DESY\,23-207\\\phantom{~}}

\hypersetup{pageanchor=false} 
\title{Vanishing of Nonlinear Tidal Love Numbers of Schwarzschild Black Holes}

\author{Massimiliano Maria Riva}
\affiliation{Deutsches Elektronen-Synchrotron DESY, Notkestr.~85, 22607 Hamburg, Germany}
\author{Luca Santoni}
\affiliation{Universit\'e Paris Cit\'e, CNRS, Astroparticule et Cosmologie, 10 Rue Alice Domon et L\'eonie Duquet, F-75013 Paris, France}
\author{Nikola Savi\'c}
\author{Filippo Vernizzi}
\affiliation{Universit\'e Paris-Saclay, CNRS, CEA, Institut de Physique Th\'eorique, 91191 Gif-sur-Yvette, France}

\begin{abstract}
\noindent
It is well known that  asymptotically flat Schwarzschild black holes in general relativity in four spacetime dimensions have vanishing induced linear tidal response. We extend this result beyond linear order for the polar sector, by solving the static nonlinear Einstein equations for the perturbations of the Schwarzschild metric and  computing the quadratic corrections to the electric-type tidal Love numbers. After explicitly performing the matching with the point-particle effective theory at leading order in the derivative expansion, we show that the Love number couplings remain zero at higher order in perturbation theory. 

\end{abstract}


%

\date{\today}
\maketitle

\section{Introduction} 

The tidal deformability of a compact object refers to its propensity  to respond when acted upon by an external long-wavelength   gravitational field. It is in general characterized in terms of complex coefficients that capture the conservative and dissipative parts of the response. The coefficients associated to the conservative response of the object are usually referred to as  \textit{Love numbers}---conceptually, one can think of  the Love numbers as the analogues of the electric polarizability of a material in electromagnetism.
The tidal response coefficients are important because they offer insights into the gravitational behavior and the body's internal structure. In the case of a neutron star, the tidal deformability is tightly related to the physics inside the object and its equation of state \cite{Baiotti:2016qnr}.
In the case of black holes, the tidal response coefficients depend on the physics at the horizon, and can be used to access and test the fundamental properties of gravity in the strong-field regime, including the existence of symmetries of the black hole perturbations \cite{Penna:2018gfx,Hui:2021vcv,Hui:2022vbh,Charalambous:2021kcz, Charalambous:2022rre, BenAchour:2022uqo, Perry:2023wmm}.   

In a binary system of compact objects, the way one body responds to the gravitational perturbation of its companion becomes more relevant in the last stages of the inspiral, influencing the waveform of the emitted gravitational waves. The tidal coefficients can be measured or constrained with gravitational-wave data. They can be used to detect binary neutron star systems \cite{LIGOScientific:2017vwq, LIGOScientific:2018hze} and have been the subject of recent searches in the LIGO-Virgo data \cite{Chia:2023tle}. 
Future observations will achieve much better accuracy and  demand high-precision calculations, such as those developed with various  schemes in \cite{Bini:2019nra,Dlapa:2021vgp,Dlapa:2022lmu,Bern:2022jvn,Jakobsen:2023hig}. The incorporation of tidal effects in these schemes will be crucial, as highlighted, e.g., in   \cite{Kalin:2020lmz,Henry:2020ski,Mougiakakos:2022sic,Heissenberg:2022tsn,Jakobsen:2023pvx}.

It is well known that isolated asymptotically flat black holes in general relativity have exactly vanishing Love numbers \cite{Damour:2009vw,Binnington:2009bb,Fang:2005qq,Kol:2011vg,Chakrabarti:2013lua,Gurlebeck:2015xpa, Porto:2016zng,LeTiec:2020spy,LeTiec:2020bos,Chia:2020yla,Charalambous:2021mea, Ivanov:2022qqt}. Quite interestingly, this result holds only in four dimensions, while higher-dimensional black holes display in general a non-vanishing conservative response~\cite{Kol:2011vg,Hui:2020xxx,Pereniguez:2021xcj,Rodriguez:2023xjd,Charalambous:2023jgq}. Most of the results in this context  have regarded so far linear perturbation theory only. However, nonlinearities are an intrinsic property of general relativity. Nonlinerities have been studied for instance in relation with quasinormal modes (see, e.g., \cite{Ioka:2007ak,Nakano:2007cj,Mitman:2022qdl,Cheung:2022rbm,Lagos:2022otp,Kehagias:2023ctr, Perrone:2023jzq,Bucciotti:2023ets}), but much less is known regarding nonlinear corrections to the tidal response of compact objects.\footnote{See \cite{Gurlebeck:2015xpa,Poisson:2020vap,Poisson:2021yau,DeLuca:2023mio} for previous works in this context. Note that our findings agree with \cite{Gurlebeck:2015xpa,Poisson:2021yau} in the particular case of axisymmetric perturbations---although we go beyond axisymmetry here. In addition, in contrast with \cite{Poisson:2020vap}, we define the nonlinear Love numbers at the level of the worldline effective theory, see section~\ref{sec:eftdefs} below. This definition has the advantage that it does not rely of any choice of coordinates.} In this work we make progress in this direction and derive quadratic corrections to the static Love numbers of Schwarzschild black holes in general relativity in four spacetime dimensions. 
{Our strategy will be to compute the response of a perturbed Schwarzschild black hole solution to an external gravitational field in the static limit and} perform the matching {up to quadratic order in the external  perturbation} with the worldline effective field theory (EFT). The latter  provides a robust framework to define the {tidal} response of compact objects \cite{Goldberger:2004jt,Goldberger:2007hy,Porto:2016pyg}.  {For simplicity, we will consider an external  field with a quadrupolar structure and even under parity transformation.}
 The two main results of our work can be summarized as follows: (i)  the {vanishing of the} linear Love numbers, defined as Wilson couplings
 of quadratic derivative operators in the worldline EFT, {is} robust against nonlinear corrections;\footnote{This is a consistency check of the natural expectation that the linear response of an object does not depend on the type of source that is used to probe it, in particular whether it has a nonlinear bulk dynamics or not.}  (ii)  the quadratic Love number couplings also vanish.

The structure of the paper is as follows. In section~\ref{sec:eftdefs}, we introduce the worldline EFT. In section~\ref{sec:GR}, we solve the  nonlinear Einstein equations up to second order in perturbation theory and in the static limit. For illustrative purposes, we will focus on the even sector only, and assume quadrupolar tidal boundary conditions at large distances for the metric perturbation.  In section~\ref{sec:matching}, we perform the matching between the EFT and the full solution in general relativity, up to second order in the external tidal field amplitude. Some details and useful technical results are collected in the appendices. In particular, {appendix \ref{app:BHPT} provides all the equations necessary for the computation of the metric solution at second order in the Regge--Wheeler gauge, while appendix \ref{App:FeynmanRules} summarizes the Feynmann rules for reference.}

\vspace{0.05in}
\noindent
\textit{Conventions.}
We use the mostly-plus signature for the metric, $(-, +, +, +)$, and work in natural units,
 $\hbar = c = 1$. We use the notation $ \kappa = \sqrt{32 \pi G} = 2\Mpl^{-1}$  and the curvature convention ${R^\rho}_{\sigma\mu\nu}=\partial_\mu\Gamma^\rho_{\nu\sigma}+\dots$ and $R_{\mu\nu}={R^\rho}_{\mu\rho\nu}$.
 We use round brackets to identify a group of totally symmetrized indices, e.g.,
		\[
			A_{(\mu|} C_{\rho} B_{|\nu)} = \frac{1}{2}\big(A_\mu C_\rho B_\nu + A_\nu C_\rho B_\mu\big) \,.
		\]
Our convention for the decomposition in spherical harmonics is 
$\Psi(t,r, \theta,\phi)= \sum_{\ell,m}  \tilde \Psi(t,r,\ell,m)Y_{\ell}^m( \theta,\phi)$. For simplicity, we will often omit the arguments on $\tilde{\Psi}$ altogether and drop  the tilde, relying on the context to discriminate between the different meanings.

\section{Worldline effective theory and Love number couplings} 
\label{sec:eftdefs}

A robust way of defining the tidal response of a compact object is in terms of the worldline EFT \cite{Goldberger:2004jt,Goldberger:2007hy,Porto:2016pyg}. By taking advantage of the separation of scales in the problem, the worldline EFT implements the idea that any object, when seen from distances much larger than its typical size,  appears in first approximation as a point source. Finite-size effects can then be consistently accounted for in terms of higher-dimensional operators localized on the object's worldline. As in any genuine EFT, they are organized as  an expansion in the number of derivatives and fields. 

Let us start from the bulk action, which we take to be the standard Einstein--Hilbert term in general relativity:
\begin{equation}
S_{\rm EH} = \int \D^4x \, \sqrt{-g} \, \frac{\MPl^2}{2} R \, .
\label{actionEH}
\end{equation}
The point-particle action is
\begin{equation}
    S_{\rm pp}= - M\int\D s=-M \int \D\tau \sqrt{-g_{\mu \nu}\frac{\D x^\mu}{\D\tau}\frac{\D x^\nu}{\D\tau}}\,.
    \label{eq:NGaction}
\end{equation}
where $M$ is the mass of the point particle, $s$ is its proper time and $\tau$ parametrizes the worldline.

To capture finite-size effects we now include derivative operators attached to the worldline. Neglecting dissipative effects \cite{Goldberger:2005cd,Goldberger:2020fot}---which are absent for the static response of nonrotating  Schwarzschild black holes---and focusing for the moment on the lowest order of the derivative expansion, the quadrupolar ($\ell=2$) Love number operators can be written as \cite{Goldberger:2004jt,Bini:2020flp,Haddad:2020que,Bern:2020uwk}
\begin{equation}
S_{\rm int } = \int \D s \sum_{n=1}^\infty \left[\lambda_{n}^{(E)}{E_{\mu_1}}^{\mu_2}\cdots{E_{\mu_{n+1}}}^{\mu_1} + \dots \right],
\label{Sint}
\end{equation}
where $E_{\mu\nu}$ is the electric (even) component of the Weyl tensor $C_{\mu\rho\nu\sigma}$, defined as
\begin{equation}
E_{\mu\nu} \equiv C_{\mu\rho\nu\sigma} u^\rho u^\sigma\, ,
\end{equation}
where $u^\mu\equiv  \D x^\mu/\D s$ is the particle's four-velocity, normalized to unity, $u^\mu u_\mu=- 1$.
Since we will focus only on the even response, in \eqref{Sint} we omitted to write explicitly operators involving the odd part of the Weyl tensor \cite{Bini:2020flp,Haddad:2020que,Bern:2020uwk}.
One can easily extend \eqref{Sint} to higher $\ell$ by introducing the multi-index operators \cite{Bern:2020uwk}
\begin{equation}
	E_{\mu_1\dots \mu_\ell}  \equiv \Ps^{\nu_3}{}_{(\mu_3}\dots \Ps^{\nu_\ell}{}_{\mu_\ell|}
	\nabla_{\nu_3}\dots \nabla_{\nu_\ell} E_{|\mu_1 \mu_2)} \, , 
\end{equation}
where $\Ps$ is the projector on the plane orthogonal to $u^\mu$, i.e.,
\begin{equation}
	\Ps^{\mu}{}_{\nu} \equiv \delta^{\mu}{}_{\nu} + u^{\mu} u_{\nu} \, .
	\label{eq:projv-def}
\end{equation}
In \eqref{Sint}, $\lambda_{n}^{(E)}$ are the (quadrupolar) Love number couplings at the $n^{\rm th}$ order in response theory. This provides an unambiguous way of defining the tidal deformability, which is independent of the choice of coordinates and the field parametrization. 
Putting all together, the EFT for the point-particle is
\begin{equation}
S_{\rm EFT} = S_{\rm EH} + S_{\rm pp} + S_{\rm int }\, .
\label{Seft}
\end{equation}
At this level, $\lambda_{n}$ are generic couplings, which will then be determined after performing the matching with the full theory.

\section{Nonlinear static deformations of Schwarzschild black holes}
\label{sec:GR}

In this section we solve the quadratic static equations for the metric perturbations of a Schwarzschild black hole in general relativity, given some suitable tidal boundary conditions at large distances.
We will denote here with $\gbRW_{\mu\nu}$ the Schwarzschild solution for the metric, $\gbRW_{\mu\nu} = \text{diag}[1-\frac{r_s}{r}, \, (1-\frac{r_s}{r})^{-1}, \, r^2,\, r^2 \sin^2\theta] $, where $r_s=2GM$,
and with $\delta g_{\mu\nu} = g_{\mu\nu}-\gbRW_{\mu\nu}$ the metric perturbation.

The quadratic equations for $\delta g_{\mu\nu}$ schematically take the form
\begin{equation}
\mathcal{D} \delta g \sim O(\delta g^2)\, ,
\label{eqschem}
\end{equation}
where $\mathcal{D}$ is a differential operator and the right-hand side is quadratic in $\delta g$. We will solve \eqref{eqschem} in perturbation theory by expanding $\delta g= \delta g^{(1)}+  \delta g^{(2)}$, where $\delta g^{(2)}\sim O((\delta g^{(1)})^2)$.  The static linearized solutions are well studied and lead to the well-known fact that the induced static response of a Schwarzschild black hole is zero, once regularity of the physical  solution is imposed at the black hole horizon~\cite{Damour:2009vw,Binnington:2009bb,Fang:2005qq,Kol:2011vg,Chakrabarti:2013lua, Gurlebeck:2015xpa,Hui:2020xxx} (see also Appendix~\ref{app:BHPT} below). Once the linear solution for $\delta g^{(1)}$ is known, the source on the right-hand side of  \eqref{eqschem} becomes fully fixed and the inhomogeneous solution to eq.~\eqref{eqschem} can be derived using standard Green's function methods. We shall stress that there are two expansion parameters in the problem: there is   $\kappa\equiv 2/\Mpl$, which controls the number of graviton field insertions, and  there is the amplitude of the external tidal field, which we will denote with $\mathcal{E}$  and which controls the nonlinear response. The two should in general be kept separate, as they appertain to different power countings in the EFT (see section~\ref{sec:matching}).

In the following we will  compute nonlinear corrections to the Love numbers by explicitly solving the second-order equations \eqref{eqschem} in some particular cases. 
As briefly reviewed in Appendix~\ref{app:BHPT}, we will parametrize the metric perturbations $\delta g_{\mu\nu}$ by distinguishing them in even (polar) and odd (axial) components, $\delta g_{\mu\nu}= \delta g_{\mu\nu}^{\text{even}} + \delta g_{\mu\nu}^{\text{odd}}$ (see eqs.~\eqref{hPMeven4D} and \eqref{hPModd4D} for the explicit expressions). 
We will assume that the external tidal field is purely even. As such, we can just focus on the even sector and set the odd perturbations $\delta g_{\mu\nu}^{\text{odd}}$ to zero: at quadratic order in perturbation theory, an external even tidal field cannot induce a parity-odd response (see Appendix~\ref{app:BHPT} for further details). 

In full generality, we shall parametrize $\delta g_{\mu\nu}^{\text{even}}$ as in eq.~\eqref{hPMeven4D}. After choosing the  Regge--Wheeler gauge \eqref{evengauge1} and solving the nonlinear $(tr)$ constraint equation, as outlined in Appendix~\ref{app:BHPT}, the expression for $\delta g_{\mu\nu}^{\text{even}}$ takes a simple diagonal  form:
\begin{equation}
\delta g_{\mu\nu}^{\text{even}} = \text{diag} \left[
\left(1-\frac{r_s}{r} \right) H_0 , \,  H_2, \, r^2 K , \,    r^2  \sin^2\theta \, K  \right] 
Y_{\ell}^m(\theta,\phi) \, ,
\label{hPMeven4Dmain}
\end{equation}
where we decomposed the field perturbations in spherical harmonics. Plugging \eqref{hPMeven4Dmain}  into the Einstein equations, one finds the following decoupled equation  for $H_0$ (see also eq.~\eqref{eqH0eveneven}):
\begin{equation}
H_0''   +\frac{2 r-r_s}{r (r-r_s)}H_0' -  \frac{ \ell(\ell+1) r (r-r_s)+r_s^2}{r^2 (r-r_s)^2} H_0 = \tilde{S}_{H_0} \, ,
\label{eqH0eveneven-main}
\end{equation}
where $\tilde{S}_{H_0}$ is fully dictated by the known linearized solution for $\delta g_{\mu\nu}^{\text{even}}$. 
Note that to write \eqref{eqH0eveneven-main} we have projected the equation for $H_0$ in real space with an $({\ell, m})$ spherical harmonic. As a result, the right-hand side of \eqref{eqH0eveneven-main} is proportional to an integral of the product of three spherical harmonics, 
\begin{equation}
\mathcal{G}^{\ell,\ell_1,\ell_2}_{ m,m_1,m_2} \equiv \int    
Y_{\ell}^{m * }(\theta,\phi)Y_{\ell_1}^{m_1}(\theta,\phi)Y_{\ell_2}^{m_2}(\theta,\phi)
 \sin\theta \D \phi \D\theta \;,
\end{equation}
which enforces the standard angular momentum selection rule $\ell=\ell_1\otimes \ell_2$. 
Given the tensor product $(\ell_1,m_1)\otimes (\ell_2,m_2)$ between two different representations of the rotation group, the resulting  total angular momentum  $\ell$  satisfies the triangular condition $\vert \ell_1-\ell_2\vert \leq \ell \leq\ell_1+\ell_2$, while the total magnetic quantum number is given by the sum $m=m_1+m_2$.
For the sake of the presentation, we will focus in the following on the case in which the external tidal field contains only a single quadrupolar harmonic, i.e., $\ell_1 = \ell_2=2$.  The analysis will be analogous with a more general tidal field and for higher harmonics.

Solving first the homogeneous linearized equation \eqref{eqH0eveneven-main} and imposing regularity at the horizon yields the following linear solution for the radial profile of $H_0$:
\begin{equation}
H_0^{(\ell=2,m)}= \mathcal{E}_m\frac{r^2}{r_s^2}\left(1- \frac{r_s}{r}  \right) \,,
\label{eqlinH020}
\end{equation}
where the amplitude $\mathcal{E}_m$ depends on the magnetic quantum number $m$.
The other components of $\delta g_{\mu\nu}^{\text{even}}$ are obtained from \eqref{eqlinH020} via the constraint equations. The linearized solutions for $H_2$ and $K$ are
\begin{equation}
H_2^{(\ell=2,m)} = \mathcal{E}_m \frac{r^2}{r_s^2} \, ,
\quad
K^{(\ell=2,m)} =   \mathcal{E}_m \frac{r^2}{r_s^2}\left(1- \frac{r_s^2}{2r^2}  \right)  \, .
\label{eqlinH2K20}
\end{equation}
Using \eqref{eqlinH020} and \eqref{eqlinH2K20}, the right-hand side of \eqref{eqH0eveneven-main} is completely fixed.  At second order, a general solution for \eqref{eqH0eveneven-main} is given by a superposition of the homogeneous solution and a  particular one. The latter can be obtained via standard Green's function methods (see Appendix~\ref{app:quadraticsolution}). One of the two integration constants for the homogeneous solution simply corresponds to a redefinition of the tidal field amplitude in \eqref{eqlinH020} and can be set to zero. The other integration constant is  chosen in such a way that the  solution at second order preserves regularity at the horizon. 
Note that, from the standard angular momentum selection rules,  an $\ell=2$ can induce at second order in perturbation theory the harmonics $\ell=0$, $2$ and $4$.   
In the following, we will focus on the quadrupole, which contributes to the leading order in the derivative expansion \eqref{Sint}.
We find the following quadratic solution for the  $(\ell=2,m)$ harmonic of $H_0$, up to quadratic order in perturbation theory:
\begin{multline}
H_0^{(\ell=2,m)}=  \frac{r^2}{r_s^2}\left(1- \frac{r_s}{r}  \right)  \Bigg[ \mathcal{E}_{m}
 \\
 - \sum_{m_1, m_2}  \mathcal{E}_{m_1} \mathcal{E}_{m_2}  \mathcal{G}^{2,2,2}_{m,m_1,m_2} \frac{ r (2 r+3 r_s)}{4r_s^2} \Bigg]   \;.
\label{eqlinH020-2}
\end{multline}
Similarly, for $H_2$ and $K$, we find 
\begin{align}
H_2^{(\ell=2,m)} & =  \frac{r^2}{r_s^2}  \left[  \mathcal{E}_m  - \sum_{m_1, m_2} \mathcal{E}_{m_1} \mathcal{E}_{m_2}  \mathcal{G}^{2,2,2}_{m,m_1,m_2} \frac{r  (4 r+r_s)}{4r_s^2} \right]  \, ,
\label{eqlinH220-2} \\
K^{(\ell=2,m)} & =  \frac{r^2}{r_s^2} \,\bigg[\mathcal{E}_m \left( 1-\frac{r_s^2}{r^2} \right)
\nonumber \\
& - \sum_{m_1, m_2} \mathcal{E}_{m_1} \mathcal{E}_{m_2}  \mathcal{G}^{2,2,2}_{m,m_1,m_2} \frac{1 }{16 } \left( \frac{2 r^2}{r_s^2} + 8   - \frac{7 r_s^2}{r^2} \right)
\bigg] .
\label{eqlinK20-2}
\end{align}
Note that the quadratic terms in $\mathcal{E}_m$ are small corrections as long as $\mathcal{E}_m r^2\ll r_s^2$. This should not surprise because the tidal field is formally divergent at large distances, and sufficiently far away perturbation theory is expected to  break down. However, in physical situations, such as in binary systems, this does not happen, because the external field acts as a growing source only on a finite region, beyond which it decays  to zero at infinity. In practice, we will perform the matching with the worldline EFT in the region $r_s\ll r \ll r_s/\sqrt{\mathcal{E}_m}$, which is sufficiently far from the black hole that the object can be treated as a point particle, but still within the range of validity of the perturbative expansion.\footnote{Note that such ``secular''-type effects are not a consequence of solving the equations in curved space. They are present also on flat space, as it can be seen for instance by formally taking in \eqref{eqlinH020-2} the limit $r_s\rightarrow0$, with $\mathcal{E}_m/r_s^2$ fixed.}

The previous results have been derived under the assumption that the external source is composed by a single quadrupolar harmonic. However,  they can be  generalized to the case  of more general tidal fields, such as a superposition of different harmonics, using the same procedure.

\section{Matching with effective theory}
\label{sec:matching}

\begin{figure}[t]
\centering
\includegraphics[scale=0.8]{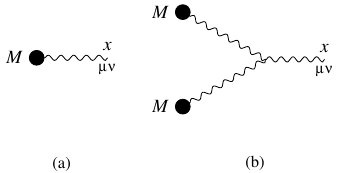}
\caption{Feynman diagrams that reconstruct the Schwarzschild metric up to order $r_s^2$.}
\label{fig:diagsrs}
\end{figure}

Given the results of section~\ref{sec:GR}, we now need to perform the matching with the worldline effective theory \eqref{Seft} and derive the Love number couplings in eq.~\eqref{Sint}.  We shall see explicitly that the matching with the calculation in general relativity can be performed with just \eqref{actionEH} and \eqref{eq:NGaction}, without turning on any of  the Love number couplings in \eqref{Sint}. 

For this computation, it is convenient to use the background field method \cite{DeWitt:1967ub,tHooft:1974toh,Abbott:1980hw}. We shall then expand the metric in eq.~\eqref{Seft} around a non-trivial background as follows
\begin{align}
	g_{\mu\nu} & = \bar{g}_{\mu\nu} + \kappa \, h_{\mu \nu} \, ,
	\label{eq:exp_BG}
\end{align}
where the background metric $\bar{g}_{\mu\nu}$ represents the external tidal field that satisfies the vacuum Einstein equations, while $h_{\mu\nu}$ parametrizes perturbative corrections in $G$ to this tidal field and, possibly, a response.

At this point, we can explicitly compute the one-point function of $h_{\mu\nu}$ induced by the  external tidal field coupled to the point particle by performing a path integral as follows:
\begin{equation}
	 \langle h_{\mu\nu}(x) \rangle = 
		\int \mathcal{D}[h] \, h_{\mu\nu}(x)
		\e^{i ( S_{\rm EFT}+ S_{\rm GF})} \, ,
	\label{eq:One-point-f}
\end{equation}
up to a normalization factor.
In the above action we have introduced the usual gauge-fixing term $S_{\rm GF}$ arising from a Faddev--Popov procedure. Since we are ultimately interested in the classical limit of the above equation, following Ref.~\cite{Goldberger:2004jt} we shall discard all diagrams with closed graviton loops. Hence, we do not need to add any ghost field. Finally, in order to maintain covariance of the final result with respect to the external metric $\bar{g}_{\mu\nu}$, we work with the following gauge-fixing action,
\begin{align}
	S_{\rm GF} & = - \int \dd^4 x \sqrt{-\bar{g}}\, \bar{g}^{\mu\nu} \bar{\Gamma}_\mu \bar{\Gamma}_\nu \, , \label{eq:GFBG}\\
	\bar{\Gamma}_\mu & \equiv \bar{g}^{\alpha\beta}\left(
		\bar{\nabla}_\alpha h_{\beta\mu} - \frac{1}{2}\bar{\nabla}_\mu h_{\alpha\beta}\right) \, .
\end{align}
Here, $\bar{\nabla}_\mu$ is the covariant derivative associated to the metric $\bar{g}_{\mu\nu}$ and $\bar{g}^{\mu\nu}$ is the inverse of the background metric.

At a practical level, we shall expand also the tidal field as 
\begin{equation}
	\bar{g}_{\mu\nu} = \eta_{\mu\nu} + H_{\mu\nu} \, .
\end{equation}
The one-point function  can then be constructed by considering all Feynman diagrams with one external $h_{\mu\nu}$. We will use  the following diagrammatic conventions: 
\begin{align*}
	\gravh & \quad \equiv \quad  h_{\mu\nu} \, , \\
	\gravH & \quad \equiv \quad H_{\mu\nu} \, , \\
	\Csoure & \quad \equiv \quad \text{point-particle source}  \, .
\end{align*}

For the comparison with section~\ref{sec:GR}, we need to compute the diagrams represented in fig.~\ref{fig:diagsrs} and \ref{fig:diagsrsE}. Their explicit expressions can be found using the Feynman rules listed in appendix~\ref{App:FeynmanRules}. We shall compute all diagrams in the rest frame of the point-particle, which means that $\tau=t$ and  the worldline is given by%
\footnote{Notice that  the normalization of $v^\mu$ with respect to the Minkowksi metric is simply $v^\mu v^\nu\eta_{\mu\nu} = -1$.} 
\begin{align}
	x^\mu = (t, 0, 0, 0) \, , & & v^\mu  = (1, 0, 0, 0) \, ,
\end{align}
where $v^\mu \equiv \frac{d x^\mu}{d \tau}$. 

The advantage of working with the background field method is that, as we mentioned, the final result for $h_{\mu\nu}$ is covariant under diffeomorphisms of the external metric $\bar{g}_{\mu\nu}$. This means that we can choose the tidal field in any convenient gauge of our choice. Hence, we choose the gauge such that $H_{\mu\nu}$ satisfies the vacuum Einstein equation on a flat background consistent with the Regge--Wheeler gauge used in section~\ref{sec:GR}. In particular, to compare with the results of that section we focus on a tidal field composed by just the harmonic $\ell=2$. Written in cartesian coordinates, this reads 
\begin{equation}
	 H_{\mu\nu}(x) = (\eta_{\mu\nu} + 2v_\mu v_\nu) A_{\alpha\beta}x^\alpha x^\beta \, ,
	\label{eq:TLO}
\end{equation}
where $A_{\mu\nu}$ is a symmetric-trace-free, purely spatial constant tensor (of mass dimension 2), i.e., $A_{\mu\nu} v^{\mu} = 0$ and $A_{\mu\nu} \eta^{\mu\nu} = 0$.
To be concrete, in spherical coordinates one has
\begin{equation}
	A_{\alpha\beta}x^\alpha x^\beta = 
	\mathcal{E}_m\frac{r^2}{r_s^2} Y_{2}^m(\theta, \phi) \, ,
	\label{eq:defA}
\end{equation}
where $r = \sqrt{x^i x^j \delta_{ij}}$ and we have chosen the amplitude $\mathcal{E}_m$ of the external tidal field in such a way as to match the notation of section~\ref{sec:GR}.

\begin{figure}[t]
\centering
\includegraphics[scale=0.8]{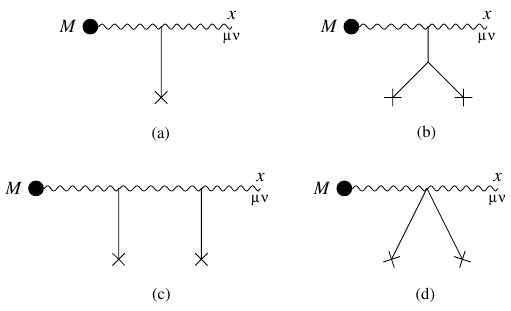}
\caption{Feynman diagrams needed for the computation of $h_{\mu\nu}$. Diagram (a) yields the order-$r_s$ correction to the linear tidal field solution. Diagrams (b), (c) and (d) represent instead order-$r_s$ corrections to the tidal source at second order in the external field amplitude.
}
\label{fig:diagsrsE}
\end{figure}

As a  sanity check, we have verified that the sum of the diagrams in figs.~\ref{fig:diagsrs} and \ref{fig:diagsrsE} satisfies the gauge condition
\begin{equation}
	\bar{\Gamma}_\mu = 0 \, ,
	\label{eq:GFexp}
\end{equation}
and  that the diagrams in fig.~\ref{fig:diagsrs} give the Schwarzschild metric up to order $G^2$ in the gauge \eqref{eq:GFexp}.\footnote{Intermediate divergences coming from diagram \ref{fig:diagsrs}{\color{blue}(b)} are handled using dimensional regularization.} This is consistent with  the well known result of, e.g., Refs.~\cite{Duff:1973zz,Goldberger:2004jt} and reproduces the background metric $\gbRW_{\mu \nu}$ in section~\ref{sec:GR}.

We can now match the result of the other diagrams to  the full-theory solution $\delta g_{\mu \nu}$ derived in section~\ref{sec:GR}. However,  while $H_{\mu\nu}$ is already in the  gauge used in section~\ref{sec:GR}, $\langle h_{\mu\nu} \rangle$ is not. Therefore, to do the comparison we
must first transform $\langle h_{\mu\nu} \rangle$ from the coordinates  $ x^\mu$ defined by the gauge condition \eqref{eq:GFexp} into the coordinate $x^\mu_{\rm RW}$ defined by the Regge--Wheeler gauge. 
The gauge transformation reads
\begin{equation}
	\kappa \langle   h^{\rm RW}_{\mu\nu} \rangle  = \kappa \langle  h_{\mu\nu} \rangle - \xi^\rho \partial_\rho\bar{g}_{\mu\nu} - 
		2\bar{g}_{\rho(\mu} \partial_{\nu)} \xi^\rho  \, ,
		\label{eq:GTexp}
\end{equation}
where $\xi^\mu = x_{\rm RW}^\mu -x^\mu $ is given below.
This allows us to define
\be
\delta \gEFT_{\mu \nu} \equiv H_{\mu \nu} +  \kappa \langle {h}^{\rm RW}_{\mu \nu} \rangle \; , 
\ee
which is now in the Regge--Wheeler gauge and can be compared to the full-theory solution $\delta g_{\mu\nu}$. 

For simplicity,  we will compare only  the $(tt)$ component, the other components of $\delta g_{\mu\nu}$ being fixed in terms of $\delta g_{tt}$   via the Einstein equations. 
Since $\langle h_{\mu\nu} \rangle$ is static, then  the gauge transformation must be time-independent. Therefore, if we focus on the $(tt)$ component, the gauge transformation simplifies to $\langle {h}^{\rm RW}_{tt} \rangle = \langle h_{tt} \rangle - \xi^i \partial_i \bar{g}_{tt}$. The derivative of the background metric is at least of order of the amplitude $A$ of the tidal field, hence, we only need to find $\xi^i$ up to order $r_s A$. Explicitly this is given by
\begin{equation}
	\xi^i = \frac{r_s}{2 r}\left( x^i + 2 r^2 A^{i j}x_{j} - A_{jk} x^j x^k x^i\right)  \, .
\end{equation}

Moreover, since $\lambda_2^{(E)}$ is independent of $m$, it is enough to perform the matching for any particular configuration of $(m, m_1, m_2)$ in order to compute its value and show that it vanishes.
For concreteness, we will consider the case where $ m_1 = 0 $ and $ m_2 = 0 $, resulting in $ m = 0 $ at second order. Performing the coordinate transformation above and projecting the result on the $(\ell=2, m=0)$ harmonic,
we find
\be
\delta \gEFT_{tt}|^{(\ell=2, m =0)} = \mathcal{E} \frac{r^2}{r_s^2} \left(1 - 2\frac{r_s}{r}\right) -\frac{1}{14} \sqrt{\frac{5}{\pi}}  \mathcal{E}^2 \frac{r^4}{r_s^4} \left(1 - \frac{r_s}{2r}\right)  \;,
\ee
where $\mathcal{E} \equiv \mathcal{E}_{m=0}$. The first term on the right-hand side, proportional to $\mathcal{E}$,  results from the  diagram \ref{fig:diagsrsE}{\color{blue}(a)} and  reproduces the $r_s/r$ correction at leading order in the tidal field \cite{Ivanov:2022hlo}. The second term on the right-hand side, proportional to $\mathcal{E}^2$, results from the last three diagrams in fig.~\ref{fig:diagsrsE}. In particular,  diagram {\color{blue}(b)} is simply the iteration of {\color{blue}(a)} due to the solution of the tidal field at order $\mathcal{E}^2$. Instead, diagrams {\color{blue}(c)} and {\color{blue}(d)} are a double insertion of the lowest-order tidal field given in eq.~\eqref{eq:TLO}. More details on the calculation of these diagrams will be provided in \cite{NonlinearLove:inprogress}.
The sum of all the four diagrams  matches the full-theory solution, $\delta g_{tt}=(1-r_s/r)H_0$, with $H_0$  given in eq.~\eqref{eqlinH020-2} for $m=m_1 = m_2 =0$, upon expanding for small $r_s/r $ and  using $\mathcal{G}^{2,2,2}_{0,0,0} = \frac17 \sqrt{\frac5{\pi}}$.

\begin{figure}[t]
\centering
\includegraphics[scale=0.8]{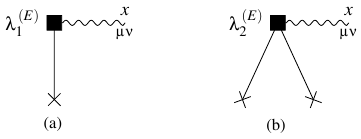}
\caption{Feynman diagrams for the (a) linear and (b) nonlinear tidal deformation. Note that these diagrams contribute at the same order in powers of the external field amplitude as the diagrams in figs.~\ref{fig:diagsrsE}{\color{blue}(a)} and \ref{fig:diagsrsE}{\color{blue}(b)}--{\color{blue}(d)}, respectively.}
\label{fig:diagT}
\end{figure}

The matching with the full theory in section~\ref{sec:GR} is obtained without the inclusion of  any of the higher dimensional operators \eqref{Sint} in the point-particle action. In other words, up to quadratic order in the external source, the couplings associated with the diagrams in fig.~\ref{fig:diagT}, which capture the induced response of the body, vanish for black holes.
%
%
 %
Note that, for the $\ell=2$ induced response, this conclusion can be reached directly from a simple dimensional analysis: the higher dimensional operators \eqref{Sint} correspond to a scaling $\sim 1/r^{\ell+1}$ in the one-point function of $h_{\mu\nu}$ which is absent in the full solution \eqref{eqlinH020-2}.
%


\section{Conclusions}

In this work, we have derived the static nonlinear response of Schwarzschild black holes in general relativity. We have explicitly solved the nonlinear Einstein equations in the static limit and up to second order in the perturbations of the Schwarzschild metric. We have then performed the matching with the worldline EFT, which provides a robust and unambiguous way of defining the tidal deformability of the object. 
At given order in powers of the external field amplitude, different types of diagrams  contribute in the EFT: there are the diagrams in fig.~\ref{fig:diagT} corresponding to the operators $E^2$ and $E^3$ in \eqref{Sint}, and there are those in  figs.~\ref{fig:diagsrs} and \ref{fig:diagsrsE} obtained from the interaction vertices in \eqref{actionEH} and \eqref{eq:NGaction}. The former  capture the true induced (linear and quadratic, respectively)  response  of the object, while the latter combine to  resum the external source. By comparing the full solution in general relativity with the EFT, we have concluded that $\lambda_1^{(E)}=\lambda_2^{(E)}=0$ in \eqref{Sint}  up to quadratic order in perturbation theory. For simplicity, we have focused on the leading order in the derivative expansion in the EFT and considered only parity-even perturbations. 
Our approach can be employed to study  higher multipoles and odd perturbations~\cite{NonlinearLove:inprogress}.

To summarize, the vanishing of the nonlinear Love numbers is a consequence of the following results: (i) at quadratic  order in perturbation theory the inhomogeneous solution is constructed from a  source (see eq.~\eqref{solHoinh}) that is made of only the linear tidal field; (ii) the point-particle EFT can be matched with the full solution without turning on Love number couplings, while  nonlinear corrections to the static solution in general relativity can be reconstructed from the EFT,  at all orders in $r_s$, via just graviton bulk nonlinearities.

The result $\lambda_1^{(E)}=\lambda_2^{(E)}=0$ was previously obtained in \cite{Poisson:2020vap} using a different approach, which relies on  harmonic coordinates and the framework of post-Newtonian theory. In contrast, our approach is not bound to the post-Newtonian expansion and is manifestly gauge invariant. Our methodology can be 
applied to prove the vanishing of other types of nonlinear Love numbers, such as those involving couplings with the gravitomagnetic field, or to compute dynamical nonlinear Love numbers beyond the static approximation; see e.g.~\cite{Poisson:2020vap,Saketh:2023bul}.
Furthermore, defining the Love numbers as Wilson coefficients of a worldline effective field theory makes it more transparent that their vanishing necessitates the existence of a nonlinear symmetry in general relativity, such as those proposed for linear fields in \cite{Hui:2021vcv,Hui:2022vbh} (see also \cite{Charalambous:2021kcz,Charalambous:2022rre,BenAchour:2022uqo}). It would be interesting to understand to what extent such  symmetries can be extended to higher orders in perturbation theory. In addition, it will be interesting to see how our conclusions change for rotating Kerr black holes,  black hole solutions in higher dimensions and different spins \cite{Hui:2020xxx, Charalambous:2021mea, Rodriguez:2023xjd,Charalambous:2023jgq,Rosen:2020crj}. We leave  all these aspects for future investigations.
\\


\subsection*{Acknowledgements}
We thank Lam Hui, Austin Joyce and Rafael Porto for useful discussions. L.S.~is supported by the French Centre National de la Recherche Scientifique (CNRS). This work was partially supported by the CNES.
M.M.R.~is partially funded by the Deutsche Forschungsgemeinschaft (DFG, German Research Foundation) under Germany's Excellence Strategy -- EXC 2121 ``Quantum Universe'' -- 390833306 and  by the ERC Consolidator Grant ``Precision Gravity: From the LHC to LISA'' provided by the European Research Council (ERC) under the European Union's H2020 research and innovation programme (grant agreement No.~817791). 
\\


\appendix

\section{Second-order perturbation theory} 
\label{app:BHPT}

In this section, we outline the derivation of the quadratic equations for the perturbations of a Schwarzschild black hole in the static limit. We shall denote the metric perturbation tensor with $\delta g_{\mu\nu}=g_{\mu\nu}-\gbRW_{\mu\nu}$, where $\gbRW_{\mu\nu}$ is the background Schwarzschild metric, and further decompose it in even (polar) and odd (axial) components as $\delta g_{\mu\nu}=\delta g_{\mu\nu}^{\text{even}}+\delta g_{\mu\nu}^{\text{odd}}$.
A general parametrization of  $\delta g_{\mu\nu}^{\text{even}}$ and $\delta g_{\mu\nu}^{\text{odd}}$ in four spacetime dimensions is given by \cite{Regge:1957td}:\footnote{Note that the definition of $h_2$ differs by a sign with respect to \cite{Regge:1957td,Franciolini:2018uyq}. In addition, the definition of $G$ differs by the subtraction of a trace, and we have reabsorbed a factor $1/(1-\frac{r_s}{r})$ in the definition of $H_2$. }
\begin{widetext}
\begin{equation}
\delta g_{\mu\nu}^{\text{even}} = \begin{pmatrix}
\left(1-\frac{r_s}{r} \right) H_0 & H_1 &  \partial_\theta \mathcal{H}_0 &   \partial_\phi  \mathcal{H}_0 \\
* & H_2 &  \partial_\theta \mathcal{H}_1 &  \partial_\phi \mathcal{H}_1 \\
* & * & r^2\left(K +  \mathcal{W} \, G  \right) &   r^2   \left(\partial_\theta  \partial_\phi - \frac{\cos\theta }{\sin\theta}\partial_\phi  \right) G    \\
* & * & * &   r^2  \sin^2\theta \left(K  - \mathcal{W}  \, G\right)
\end{pmatrix}  \, ,
\label{hPMeven4D}
\end{equation}
\begin{equation}
\delta g_{\mu\nu}^{\text{odd}} = \begin{pmatrix}
0 & 0 & - \frac{1}{\sin\theta} \partial_\phi h_0 & \sin\theta \partial_\theta h_0 \\
* & 0 & - \frac{1}{\sin\theta} \partial_\phi h_1 &  \sin\theta \partial_\theta h_1 \\
* & * & -\frac{1}{\sin\theta}\left( \partial_{\theta}\partial_{\phi} - \frac{\cos\theta}{\sin\theta}\partial_\phi \right) h_2 & \frac{1}{2} \sin\theta\left( \partial_{\theta}^2 - \frac{\cos\theta}{\sin\theta}\partial_\theta - \frac{1}{\sin^2\theta} \partial_\phi^2 \right) h_2 \\
* & * & * &   \sin\theta\left( \partial_{\theta}\partial_{\phi} - \frac{\cos\theta}{\sin\theta}\partial_\phi \right) h_2
\end{pmatrix}   ,
\label{hPModd4D}
\end{equation}
\end{widetext}
where the asterisks denote symmetric components and where we introduced the differential  operator $\mathcal{W} \equiv \frac{1}{2}( \partial_\theta^2 -  \frac{\cos\theta}{\sin\theta}\partial_\theta -\frac{1}{\sin^2\theta}\partial_\phi^2 )$. Each component of $\delta g_{\mu\nu}^{\text{even}}$ and $\delta g_{\mu\nu}^{\text{odd}}$ can be further decomposed in spherical harmonics as, for instance, $H_0= \sum_{\ell , m} H_0 Y_{\ell}^m( \theta,\phi)$, where   $Y_{\ell}^m(\theta,\phi)$ are normalized as $\int \D\Omega \, Y_{\ell}^{m*}(\theta,\phi) Y_{\ell'}^{m'}(\theta,\phi) =\delta_{\ell\ell'}\delta^{mm'}$.
Since $\delta g_{\mu\nu}^{\text{even}}$ and $\delta g_{\mu\nu}^{\text{odd}}$ have opposite transformation rules under a parity transformation,  the spherical symmetry of the background $\bar{g}_{\mu\nu}$ ensures that $\delta g_{\mu\nu}^{\text{even}}$ and $\delta g_{\mu\nu}^{\text{odd}}$ do not couple at the level of the linearized equations of motion. Mixing will appear starting from quadratic order.

\subsection{Quadratic solution from even tidal field} 
\label{app:quadraticsolution}

In this section, we derive the even quadratic equations for the metric perturbations in the static regime  and solve them under the assumption of a purely even tidal field at large distances. Without an odd tidal field, since the linear odd static solution is divergent at the horizon,\footnote{This is equivalent to saying that the linear Love numbers vanish.} we can set  $\delta g_{\mu\nu}^{\text{odd}}$ to zero altogether and just focus on the even perturbations $\delta g_{\mu\nu}^{\text{even}}$.

\paragraph*{Gauge fixing.} At each order in perturbation theory, we choose to fix the {Regge--Wheeler} gauge  as follows  \cite{Regge:1957td,Nakano:2007cj,Brizuela:2009qd}: 
\begin{equation}
\mathcal{H}_0 = \mathcal{H}_1 = G=0\, .
\label{evengauge1}
\end{equation}
Since this is a complete gauge fixing, it can be performed directly in the action without losing any constraints \cite{Motohashi:2016prk}.

\paragraph*{Constraint equation.} With the gauge choice \eqref{evengauge1}, the only off-diagonal metric component in $\delta g_{\mu\nu}^{\text{even}}$ is thus  $H_1$, which is a constrained variable.  It is not hard to see that, in the static limit and in the absence of odd perturbations, 
\begin{equation}
H_1 =0 \, ,
\label{solcinstraintH1-1}
\end{equation}
at each order in perturbation theory. This follows from solving  $G_{tr}=0$, where $G_{\mu\nu}= R_{\mu\nu}-\frac{1}{2}g_{\mu\nu}R$ is the Einstein tensor. In fact, by construction, $G_{tr}$  is at each order proportional to (derivatives of) $H_1$, i.e., it vanishes when $H_1$ vanishes.  
As a result, with the gauge choice \eqref{evengauge1} and the nonlinear solution \eqref{solcinstraintH1-1} in the static limit, the metric perturbation $\delta g_{\mu\nu}^{\text{even}}$ boils down to the diagonal form: $\delta g_{\mu\nu}^{\text{even}}=\text{diag} \left[ (1-\frac{r_s}{r} ) H_0, H_2, r^2 {K},   r^2  \sin^2\theta {K} \right]$ \cite{Hinderer:2007mb}. Next, we plug this into the Einstein--Hilbert action, which we expand up to cubic order in the perturbations $H_0$, $H_2$ and  ${K}$. Taking then the variation with respect to each of the three metric components, we can write down the quadratic equations for $H_0$, $H_2$ and  ${K}$. Two of them will lead to constraints, while only one will give the static equation for the physical (even) degree of freedom. To make this manifest, it is first convenient to perform the following field redefinition,
\begin{equation}
H_2 \equiv \psi  + \frac{r^2}{r-r_s} { K}' + \frac{r^2 }{2 (r-r_s)^2} \left[ 2- \frac{3r_s}{r} + \Delta_{S^2} \right]  { K} \, ,
\end{equation}
where $\Delta_{S^2}= \partial_\theta^2 +  \frac{\cos\theta}{\sin\theta}\partial_\theta  + \frac{1}{\sin^2\theta}\partial_\phi^2$ is the spherical Laplacian on the $2$-sphere, defined with line element $\D \Omega^2\equiv \D \theta^2+\sin^2\theta\,  \D\phi^2$ and satisfying $\Delta_{S^2}   { K}= -\ell(\ell+1) { K}$.
The resulting equations are 
\begin{widetext}
\begin{equation}
\psi ' + \frac{\left(\ell^2+\ell+2\right) r+2 r_s}{2 r (r-r_s)} \psi - \frac{\ell (\ell+1) r \left[\left(\ell^2+\ell-2\right) r+3 r_s\right]}{4 (r-r_s)^3} K   = \frac{r^2 }{r-r_s} \int\D\Omega \, Y_{\ell}^{m*}(\theta,\phi)  S_1(r,\theta,\phi) \, ,
\label{eqeveneven1}
\end{equation}
\begin{equation}
K' - 2\left(1-\frac{r_s}{r}\right)H_0'  - \frac{\ell(\ell+1)+1}{r-r_s}K  -\frac{\ell(\ell+1)}{r_s}H_0 - \frac{2}{r}\left(1-\frac{r}{r_s}\right)\psi 
= -\frac{4 r }{\Mpl^2 r_s(r-r_s) } \int \D\Omega  \, Y_{\ell}^{m*}(\theta,\phi) \csc \theta S_2(r,\theta ,\phi ) \, ,
\label{eqeveneven2}
\end{equation}
\begin{multline}
\psi' +\frac{r [r_s-\ell (\ell+1) r] }{r_s (r-r_s)} H_0' +\frac{r [r_s-\ell (\ell+1) r] }{2 (r-r_s)^2}K'
-\frac{\ell(\ell+1)r(r-r_s)-r_s(2r+r_s)}{r r_s (r-r_s)} \psi
\\
+ \frac{\ell (\ell+1) r (r-2 r_s)}{r_s (r-r_s)^2}H_0
-\frac{r\left[\ell(\ell+1) \left(\ell^2+\ell-5\right) r r_s-(\ell-1) \ell (\ell+1) (\ell+2) r^2+(2 \ell r_s+r_s)^2\right]}{2 r_s (r-r_s)^3} K
\\ = - \frac{4 r^3   }{\Mpl^2 r_s (r-r_s)^2 }   \int \D\Omega  \, Y_{\ell}^{m*}(\theta,\phi)  \csc \theta S_3(r,\theta ,\phi ) \, ,
\label{eqeveneven3}
\end{multline}
\end{widetext}
where $S_1$, $S_2$ and $S_3$ are source terms, quadratic in the fields---which we will not write explicitly. The goal is to solve \eqref{eqeveneven1}-\eqref{eqeveneven3} in perturbation theory. After straightforward manipulations, one finds that the field components $H_2$ and ${ K}$ can be solved algebraically for in terms of $H_0$ and derivatives thereof. Hence, the problem reduces to solving the $H_0$'s equation of motion, which, after some massaging of \eqref{eqeveneven1}-\eqref{eqeveneven3},  is found to be
\begin{equation}
H_0''   +\frac{2 r-r_s}{r (r-r_s)}H_0' -  \frac{ \ell(\ell+1) r (r-r_s)+r_s^2}{r^2 (r-r_s)^2} H_0 = \tilde{S}_{H_0} \, ,
\label{eqH0eveneven}
\end{equation}
where $\tilde{S}_{H_0}$ is a linear combination of (derivatives of) the source terms in \eqref{eqeveneven1}-\eqref{eqeveneven3}.

The homogeneous part of \eqref{eqH0eveneven} is a (degenerate) hypergeometric equation, which can thus be solved in closed form.
To bring it in standard hypergeometric form, it is convenient to perform the following field redefinition,
\begin{equation}
H_0(r(x)) = x^{\ell+1}(1-x)^{-1} u(x) \, ,
\qquad
x\equiv \frac{r_s}{r} \, .
\label{fieldredhyper}
\end{equation}
Using \eqref{fieldredhyper}, the homogeneous equation takes on the form
\begin{equation}
x(1-x) u''(x) + [\mathit{c}- (\mathit{a}+\mathit{b}+1)x ] u'(x) -\mathit{a} \mathit{b} u(x)=0 \, ,
\end{equation}
with parameters
\begin{equation}
a= \ell-1\, ,
\qquad
b= \ell+1 \, ,
\qquad
c= 2\ell+2 \, ,
\end{equation}
satisfying the relation $c-a-b=2$. The two linearly independent solutions are\footnote{This case corresponds to line 20 of the table in Sec.~2.2.2 of \cite{Bateman:100233}, with $m=\ell-2$, $n=\ell$ and $l=2$. The two independent solutions can be found in eqs.~2.9(1) and 2.9(13) of \cite{Bateman:100233}. There is a typo in the case 20 of the table in Sec.~2.2.2: the ``$u_2$'' should be instead ``$u_4$''. }
\begin{align}
u_1(x) & = \hypergeom{2}{1}(\ell-1,  \ell+1,  2\ell+2; x)\, ,
\\
u_4(x) & = (-x)^{-\ell-1}\hypergeom{2}{1}(-\ell,  \ell+1,  3; x^{-1}) \, .
\end{align}
Since the first argument of $u_4$ is a non-positive integer and the third argument is a positive number, we can use the formula \cite{Hui:2020xxx}
\begin{equation}
\begin{split}
(-x)^{-\ell-1}\hypergeom{2}{1}  & (-\ell,  \ell+1,  3; x^{-1}) =
		\\
& = (-x)^{-\ell-1 }\sum_{n=0}^{\ell} \frac{(-\ell)_n(\ell+1)_n}{(3)_n n!}x^{-n} \, ,
\end{split}
\label{case3appreg}
\end{equation}
where $(\cdot)_n$ is the Pochhammer symbol.
Notice that only this second solution leads to a $H_0$ that is regular at $x=1$ ($u_1$ contains instead  a logarithmic divergence). In particular, $H_0$ constructed out of $u_4$ is a finite polynomial with only positive powers of $r$---we recover in other words the well-known fact that Love numbers of black holes in four spacetime dimensions vanish.

In terms of $H_0$, the two linearly independent solutions read
\begin{align}
H_0^{(1)}(r) & = \frac{\left( \frac{r_s}{r}\right)^{\ell+1}}{1- \frac{r_s}{r}} \hypergeom{2}{1}(\ell-1,  \ell+1,  2\ell+2;  \tfrac{r_s}{r})\, ,
\label{H01} \\
H_0^{(4)}(r) & = \frac{ \left(-1\right)^{\ell+1}}{\left(1- \frac{r_s}{r}\right)} \hypergeom{2}{1}(-\ell,  \ell+1,  3; \tfrac{r}{r_s}) \, .
\label{H04} \\ \nonumber
\end{align}
The solution to the inhomogeneous solution \eqref{eqH0eveneven} is
\begin{equation}
H_0(r) = \int_{r_s}^\infty G(r,r') \tilde{S}_{H_0}(r') \D r' \, ,
\label{solHoinh}
\end{equation}
where the Green's function satisfies
\begin{equation}
\left[ \partial_r^2   +\frac{2 r-r_s}{r (r-r_s)}\partial_r -  \frac{ \ell(\ell+1) r (r-r_s)+r_s^2}{r^2 (r-r_s)^2}\right] G(r,r') = \delta(r-r')\, .
\label{GFeq}
\end{equation}
For $r\neq r'$, the most general solution for $G(r,r')$ that is regular at the horizon and is continuous across $r=r'$ is given by the combination 
\begin{multline}
G(r,r') = \frac{1}{W(r')} \Big[ H_0^{(1)}(r)  H_0^{(4)}(r') \theta(r-r') 
\\
+  H_0^{(1)}(r')  H_0^{(4)}(r) \theta(r'-r) \Big] \, ,
\end{multline}
where  $H_0^{(1)}(r)$ and $H_0^{(4)}(r)$ can be read off from eqs.~\eqref{H01} and \eqref{H04}, and where $W$ is the Wronskian, 
\begin{equation}
\begin{split}
W(r) &=   H_0^{(4)}(r) \partial_r H_0^{(1)}(r)  -  H_0^{(1)}(r) \partial_r H_0^{(4)}(r) 
\\
&	= W_0 \frac{r_s}{r} \left(\frac{r}{r_s} -1 \right)^{-1} \, .
\end{split}
\label{solA}
\end{equation}
$W_0$ is  an   $\ell$-dependent constant, which can be written in closed form as
\begin{widetext}
\begin{multline}
W_0 = \frac{1}{3 r_s} (-1)^{-\ell} 2^{-\ell-2} \bigg\{ 3 (\ell-1)  \hypergeom{2}{1}(-\ell,\ell+1;3;2) \hypergeom{2}{1}\left(\ell,\ell+2,2 \ell+3;\tfrac{1}{2}\right)
\\
-4 (\ell+1) \hypergeom{2}{1}\left(\ell-1,\ell+1,2 \ell+2;\tfrac{1}{2}\right) \left[2 \ell \hypergeom{2}{1}(1-\ell,\ell+2,4;2)-3 \hypergeom{2}{1}(-\ell,\ell+1,3;2)\right]\bigg\} \, .
\end{multline}
\end{widetext}
All in all, the inhomogeneous solution that is regular at the horizon can be written as
\begin{multline}
H_0(r) = \frac{1}{W_0 r_s^2} \bigg[  H_0^{(1)}(r)  \int_{r_s}^r   H_0^{(4)}(r')  (r'^2-r_s r')\tilde{S}_{H_0}(r')  \D r' 
\\
+ H_0^{(4)}(r) \int_{r}^{r_0}   H_0^{(1)}(r')  (r'^2-r_s r')\tilde{S}_{H_0}(r')  \D r'  \bigg]  \, .
\label{app:H0regh0}
\end{multline}
In \eqref{app:H0regh0} we introduced an arbitrary radius $r_0$. This is completely immaterial as the integral evaluated at $r_0$ can always be reabsorbed by a redefinition of the integration constant of the homogeneous solution that is regular at the horizon.

\section{List of Feynman rules}
\label{App:FeynmanRules}

In this  appendix, we  list the Feynman rules used in the main text  for the EFT matching computation:
\begin{widetext}
\begin{align}
\proph & = \frac{i}{k^2 - i0^+} P_{\mu\nu\rho\sigma} \, ,  \\
\cubic & = i \kappa (2\pi)^4\delta^{(4)}(k_1 + k_2 + k_3) \mathcal{V}_3^{\alpha_1 \beta_1 \alpha_2 \beta_2 \alpha_3 \beta_3}  \, ,\\
\cubicback & = i H_{\alpha_1 \beta_1}(-k_2 -k_3) V_3^{\alpha_1 \beta_1}{}_{\alpha_2 \beta_2 \alpha_3 \beta_3} \, , \\
\quarticback & = i \int\!\frac{\dd^4 q}{(2\pi)^4} H_{\alpha_1 \beta_1}(q)H_{\alpha_2 \beta_2}(-k_2 -k_3-q) V_4^{\alpha_1 \beta_1\alpha_2 \beta_2}{}_{ \alpha_3 \beta_3\alpha_4 \beta_4} \, , \\
\Source & = i \kappa\frac{M}{2}\int \dd \tau \, \e^{- i k\cdot x(\tau)} v^\mu v^\nu \, ,
\label{eq:FRS}
\end{align}
\end{widetext}
where the cubic vertex $\mathcal{V}_3$ is obtained from expanding the Einstein--Hilbert action up to cubic order in $h_{\mu\nu}$. The cubic and quartic vertices $V_3$ and $V_4$ are obtained by expanding eqs.~\eqref{actionEH} and \eqref{eq:GFBG} up to quadratic order in both $h_{\mu\nu}$ and $H_{\mu\nu}$. 
Their tensorial structures are handled using the xAct package for Mathematica \cite{xAct}.

\bibliographystyle{utphys}  
\bibliography{biblio}


\end{document}